\documentclass[a4paper, 10pt, conference]{ieeeconf}      
\usepackage{FG2026}

\FGfinalcopy 

\IEEEoverridecommandlockouts                              
\usepackage{graphics} 
\usepackage{epsfig} 
\usepackage{amsmath} 
\usepackage{booktabs}
\usepackage{soul}
\usepackage{multirow}

\usepackage{fancyhdr}
\def\FGPaperID{434} 

\title{\LARGE \bf
An Algorithm for On-Sensor Agnostic Detection of Changes in Human Activity for Ultra-Low-Power Applications
}

\author{\parbox{16cm}{\centering
   {\large Sara Rimoldi, Arianna De Vecchi, Hazem Hesham Yousef Shalby, Federica Villa\\}
   {\normalsize
   Dipartimento di Elettronica, Informazione e Bioingegneria (DEIB), Politecnico di Milano, Milan, Italy\\}}
}

\begin{document}

\ifFGfinal
\thispagestyle{empty}
\pagestyle{empty}
\else
\author{Anonymous FG2026 submission\\ Paper ID \FGPaperID \\}
\pagestyle{plain}
\fi
\maketitle

\thispagestyle{fancy}

\begin{abstract}
Wearable devices running Human Activity Recognition~(HAR) on Inertial Measurement Units~(IMUs) waste energy by performing continuous classification for each window, even during long periods of unchanged activity.
We address this with a lightweight change-detection gate: a non-parametric algorithm based on dynamic template matching that runs continuously at only approximately 16\,kFLOPs per step, requires no offline training, and does not need prior definition of target activity classes.
The gate invokes the full HAR network only when it detects an activity change, reducing the computational load by over~67\% in realistic monitoring settings.
The algorithm is evaluated on smart glasses, smartwatch, and smartphone data, requiring only a brief device-specific calibration phase. The gate achieves 98\% sensitivity on UCA-EHAR, ensuring no genuine activity transition is missed, while 75\% specificity keeps unnecessary HAR invocations low.
Results on WISDM are 97\% sensitivity and 76\% specificity, demonstrating robustness and flexibility to various settings.
\end{abstract}


\section{Introduction}
\label{sec:intro}

Over the last few years, on-the-edge Human Activity Recognition (HAR) based on the use of Inertial Measurement Units (IMUs) has emerged as a topic of interest due to its inherent advantages of privacy, low latency, and reduced power consumption~\cite{Yin2024}. 

The latter, however, remains a critical bottleneck. Wearables operate on small batteries and must sustain continuous sensing for hours or days. This represents an inefficient use of system resources, as in realistic settings human activity does not change continuously, thus not requiring constant classification.
As such, deploying continuous, always-on HAR solutions for the classification of prolonged repetitive activities incurs significant and avoidable power consumption~\cite{DeVecchi2025, Shalby2025}.

\begin{figure}[b]
\centering\includegraphics[width=1\linewidth]{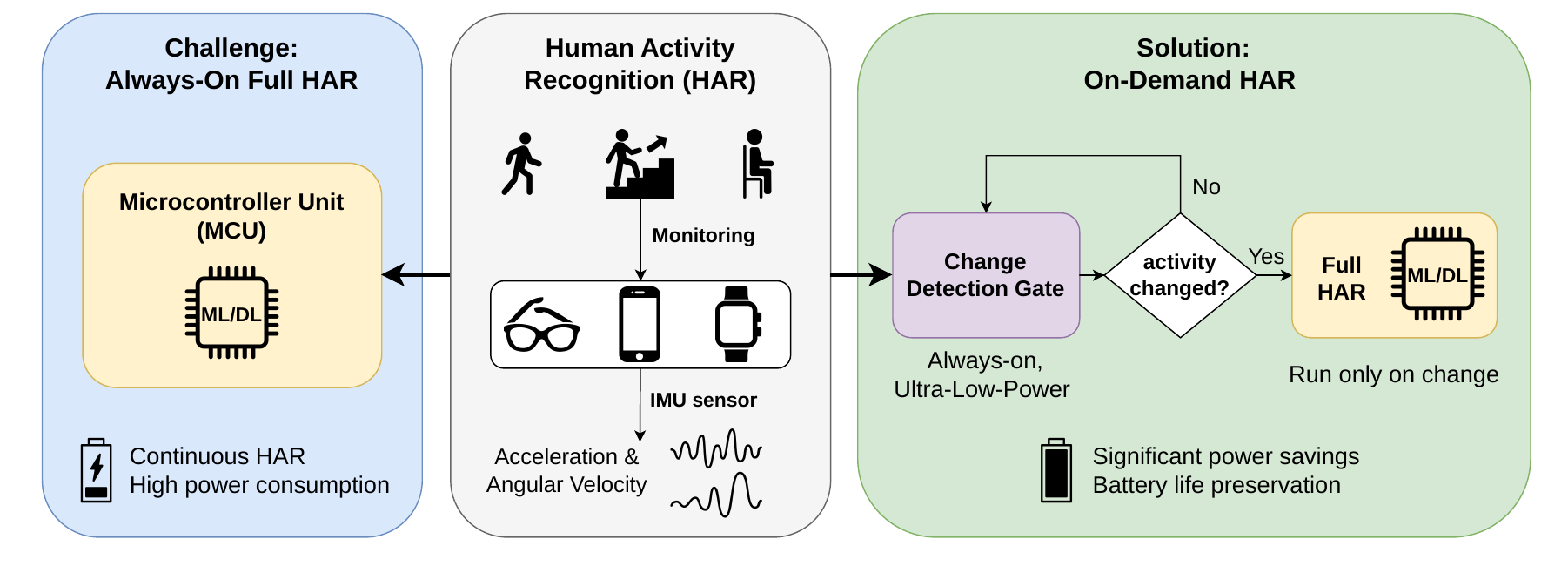}
    \caption{Proposed partitioned computation approach, which aims to avoid always-on HAR. MCU: Microcontroller Unit; HAR: Human Activity Recognition.}
    \label{fig:diag_general}
\end{figure}

We propose a lightweight change-detection gate to be executed before the HAR network, which is hence performed only when a change is detected, as described in Fig.~\ref{fig:diag_general}. In particular, the algorithm employs inertial data to build a template specific to the current activity starting from feature distributions and compares it with a stored reference template. If, with respect to a predefined threshold, the templates differ, the activity has changed, the reference is updated, and the full HAR pipeline is triggered.
This approach does not require training, but only a brief one-time threshold calibration, which requires only a few minutes of labelled transitions and does not demand storing raw sensor data.
The gate requires only 4.53\,kB and approximately 16\,kFLOPs per step, which are negligible when compared to the continuous execution of any HAR network, e.g., $853$\,kFLOPs in~\cite{11007628}.

We validate the approach on UCA-EHAR~\cite{Novac2022} and WISDM~\cite{Weiss2019}, comprising data from different wearable and mobile devices, specifically smart glasses, a smartwatch, and a smartphone, while only requiring a brief device-specific tuning. The gate achieves an average sensitivity of 98\% on UCA-EHAR and of 97\% on WISDM, thus ensuring every type of activity transition triggers an HAR update. The 75\% and 76\% specificity values on UCA-EHAR and on WISDM, respectively, prevent unnecessary invocations, while also allowing the gate to adapt to the current activity.

\section{Related Works}
\label{sec:previous}

Although HAR architectures have been successfully deployed on resource-constrained devices, including microcontrollers (MCUs)~\cite{11007628} and smart sensors~\cite{DeVecchi2025, Ronco2022}, some issues remain unsolved. In particular, even if network quantization and pruning substantially reduce model size and computational load, these methods still require to perform continuous inference and therefore remain unsuitable for power-constrained settings, as inference must be executed on every window, even when no activity change has occurred. 
Hierarchical approaches, such as Dendron~\cite{11007628}, as well as early-exits strategies~\cite{montello_survey_2025}, are able to mitigate this drawback by lowering the computational cost, but do not address the issue of avoidable activations. 


To solve this, \cite{Shalby2025} presents a gating approach in which a lightweight algorithm is continuously executed on a smart sensor, while the HAR network infers on data only when a specific condition is identified. Specifically, the work suggests as an example that HAR activation could be triggered once the device detects it is being worn. However, this method still requires a dataset and the corresponding training phase in order to deploy the gate algorithm.
In order to improve this, a Change-Point Detection~(CPD) algorithm could be introduced, providing a framework for identifying distributional shifts in sequential data~\cite{Aminikhanghahi2017}.
In the literature, a critical limitation of CPD approaches such as CUSUM~\cite{Buroni2024} is their frequent dependence on the a priori definition of rigid statistical parameters, such as sensitivity ratios or fixed thresholds, which severely restrict the algorithm's generalisation capabilities. On the other hand, non-parametric approaches often require complex mathematical steps that are computationally prohibitive for real-time, online execution \cite{Aminikhanghahi2017, Buroni2024}.

This paper aims to fill the gap in the CPD literature, providing a lightweight algorithm, able to run in real-time and performing as a gate for an HAR network.
To the best of our knowledge, no prior method simultaneously satisfies:
(i)~no offline training, apart from a brief once-only calibration,
(ii)~no distributional assumptions on the sensor signal, and
(iii)~a compute and memory budget compatible with severely resource-constrained wearable platforms, including smart sensors~\cite{STMicroelectronics_ISPU}.

\label{sec:methods}
\section{Methods}

\begin{figure*}[t!]
    \centering
    \includegraphics[width=1.0\linewidth]{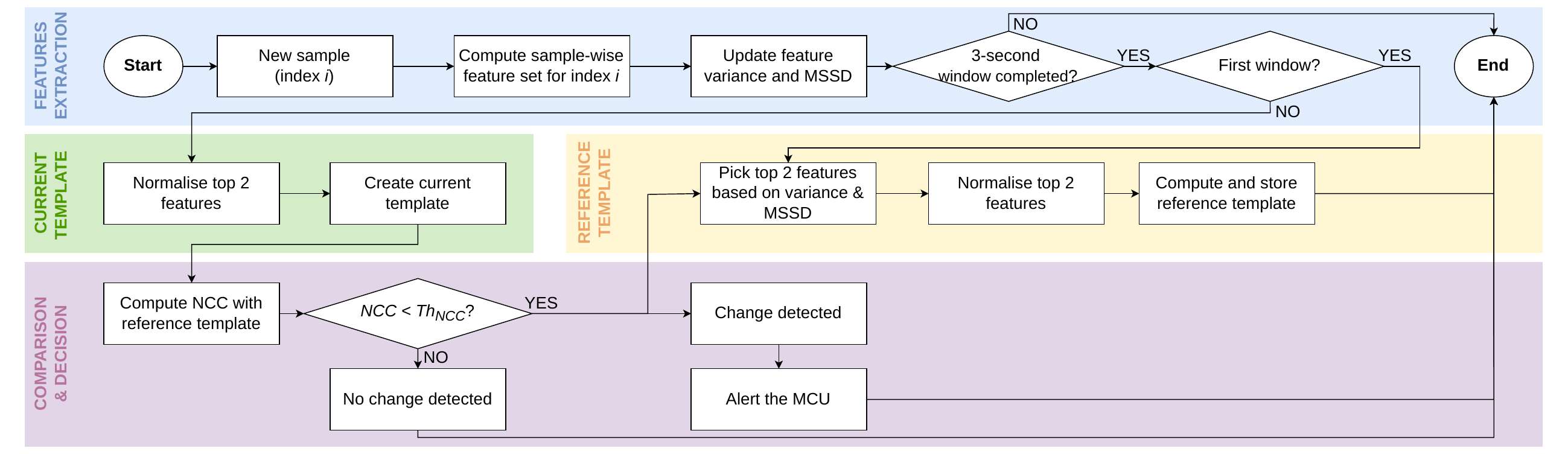}
    \caption{High-level workflow of the main phases of the algorithm.}
    \label{fig:workflow_algo}
\end{figure*}

The algorithm is composed of four main phases, which are executed as shown in Fig.~\ref{fig:workflow_algo}: \textbf{(A)}~feature extraction, \textbf{(B)}~reference template creation, \textbf{(C)}~current template creation, and \textbf{(D)}~comparison and decision.


\subsection{Feature Extraction}
\label{sub:features}
The feature extraction process is iterated over a window with $W$ samples.
At each time instant $i$ in the window, the IMU provides accelerations $(a_x, a_y, a_z)[i]$ and angular velocities $(\omega_x, \omega_y, \omega_z)[i]$, from which a feature vector $\mathbf{f}[i] \in \mathbf{R}^{12}$ is computed in a sample-wise manner, thus without buffering raw data. Beyond restricting the memory footprint to stored feature values, this design also guarantees real-time execution by avoiding complex window-wise computations that could otherwise introduce processing bottlenecks and delay sample acquisition.
The features and the equations used to compute them are presented in Table~\ref{tab:features_computation}.

The 12 features were selected among the most commonly used ones in state-of-the-art HAR approaches~\cite{Yin2024}, and span five categories: magnitude (Euclidean norms of $\mathbf{a}$ and $\boldsymbol{\omega}$), rate of change (first-order derivatives of $a_x$, $\omega_y$, $\|\boldsymbol{\omega}\|$), orientation (gravity components $g_x, g_y, g_z$ via Euler angles), mean-crossing amplitudes of $\omega_x$, $\omega_y$ via recursive accumulators, and peak-to-peak amplitudes of $\|\mathbf{a}\|$, $\omega_x$ via recursive min/max tracking.
By adopting features widely used in state-of-the-art HAR approaches~\cite{s18124189}, the same computed values can be used both for transition detection and downstream classification, without additional extraction steps.
\begin{table*}[t!]
    \caption{Feature equations for sample $i$. Subscript $u$ denotes acceleration $a$ or angular velocity $\omega$; components are along $X$, $Y$, $Z$. Features were extracted from a subset of axes, as detailed in Subsection~\ref{sub:features} and in the column \#Features.}
    \label{tab:features_computation}
    \begin{center}
    \renewcommand{\arraystretch}{1.3}
    \scriptsize
    \begin{tabular}{llc}
        \toprule
        \textbf{Feature} & \textbf{Equation} & \textbf{\#Features}  \\
        \midrule
        Euclidean norm & $norm_u[i] = \sqrt{(u_x[i])^2 + (u_y[i])^2 + (u_z[i])^2}$  & 2\\
        First-order derivative & $der_u[i] = u[i] - u[i-1]$ & 3\\
        Gravity component x & $g_x[i] = \sin (\theta[i])$ & 1 \\
        Gravity component y & $g_y[i] = \cos( \theta[i] )\sin(\phi[i])$ & 1\\
        Gravity component z & $g_z[i] = \cos (\theta [i])\cos( \phi[i])$ &1\\
        Mean-crossing rate & $mc_u[i] = mc_u[i-1] \pm |u[i] - avg_u[i-1]|$ if $cross_{\pm}[i]$; else $mc_u[i-1] \cdot \gamma_{mc}$ &2 \\
        Peak-to-peak amplitude & $p2p_u[i] = max_u[i] - min_u[i]$ & 2\\
        \midrule
        &&\textbf{Total: 12}\\
        \bottomrule
        
        \toprule
        \multicolumn{2}{c}{\textbf{Auxiliary Variables}} \\
        \midrule
        \textbf{Feature} & \textbf{Equation}\\
        \midrule
        Roll angle & $\phi[i] = \mathrm{arctan2}(a_y[i], a_z[i])$ \\
        Pitch angle & $\theta[i] = \arcsin\left(\frac{a_x[i]}{\sqrt{(a_x[i])^2 + (a_y[i])^2 + (a_z[i])^2}}\right)$ \\
        Running mean & $avg_u[i] = \frac{1}{i+1} \sum_{j=0}^{i} u[j]$ \\
        Running variance & $var[i] = \frac{1}{{i+1}}\sum_{j=0}^{i}  u[j]^{2} - (avg_u[i])^{2}$ \\
        Upward crossing & $cross_+[i] = 1$ if $(u[i-1] < avg_u[i-1] - hyst\ \land\ u[i] \ge avg_u[i-1] + hyst)$; else $0$ \\
        Downward crossing & $cross_-[i] = 1$ if $(u[i-1] > avg_u[i-1] + hyst\ \land\ u[i] \le avg_u[i-1] - hyst)$; else $0$ \\
        Running maximum & $max_u[i] = u[i]$ if $u[i] > max_u[i-1]$; else $avg_u[i-1] + \gamma_{p2p} \cdot(max_u[i-1] - avg_u[i-1])$ \\
        Running minimum & $min_u[i] = u[i]$ if $u[i] < min_u[i-1]$; else $avg_u[i-1] - \gamma_{p2p} \cdot (avg_u[i-1] - min_u[i-1])$ \\
        Recursive decay factors &  $\gamma_{mc}=0.8$; $\gamma_{p2p}=0.7$ \\
        \bottomrule
    \end{tabular}
    \end{center}
\end{table*}

\subsection{Reference Template Creation}
\label{sec:template}

In order to select the two most informative features that best capture the signal in the selected window, two statistical dispersion parameters, the variance and the Mean of the Squared Successive Differences (MSSD), are iteratively updated for each feature.
Denoting the value of feature $f$ at sample $i$ as $x_f[i]$, the Welford updating rule~\cite{desa2019} to retrieve the recursive variance is defined, as in (\ref{eq:variance_welford}):

\begin{equation}
    \label{eq:variance_welford}
    var_f[i] = \frac{M_{2,f}[i]}{i},
\end{equation}
where $M_{2,f}$ is the running sum of square deviations for $f$, computed using (\ref{eq:variance_params_update}). No variance value is associated with the first sample.

\begin{equation}
    \label{eq:variance_params_update}
    \begin{cases}
    \delta_f[i] = x_f[i] - \mu_f[i-1] \\
    \mu_f[i] = \mu_f[i-1] + \frac{1}{i+1}\cdot\delta_f[i] \\
    M_{2,f}[i] = M_{2,f}[i-1] + \delta_f[i] \cdot (x_f[i] - \mu_f[i]) \\
    \end{cases}
\end{equation}

In the equation, $\mu_f[i]$ is the running average of feature $f$, and $\delta_f[i]$ is its incremental correction.

The MSSD is a statistical measure that quantifies the short-term variability of a signal, often employed to quantify randomness in patterns \cite{Li2018}.
The computation of the MSSD (except for the first sample) is defined in (\ref{eq:mssd_formula}).
\begin{equation}
    MSSD_f[i] = \frac{1}{i+1} \sum_{i=1}^{N_i-1} \left( x_f[i] - x_f[i-1] \right)^2
    \label{eq:mssd_formula}
\end{equation}

High variance reflects sustained feature amplitude, while low MSSD indicates periodic, repetitive features. Together, the two statistics are normalised by the squared Root Mean Square (RMS) of feature $f$ to account for differing dynamic ranges~\cite{Li2018}.
Therefore, a score based on variance maximisation and MSSD minimisation is assigned to each feature, and the two best ones are selected and employed for template creation.

$f_1$ and $f_2$ denote the two selected features, which are then normalised in a [0;1] range employing (\ref{eq:rescaling}). $max(f)$ and $min(f)$ are the window-wise extrema of feature $f$, and are updated whenever a transition is detected.
\begin{equation}
    \hat{x_f}[i] = \frac{x_f[i] - \min(f)}{\max(f) - \min(f)}
    \label{eq:rescaling}
\end{equation}
The normalisation step ensures that the features are constrained in a finite interval, despite originally having different amplitude ranges.
The normalised pair $(\hat{x_f}_1[i], \hat{x_f}_2[i])$ is then accumulated into a $10 \times 10$ template, in a procedure similar to the one presented in~\cite{Chowdhary2023}.
In particular, the template is created as a normalised 2-D histogram, where the first normalised feature $\hat{f}_1$ is placed on the horizontal axis, and the second $\hat{f}_2$ is placed along the vertical axis. 

The histogram's bin counts are further normalised by the total number of samples to obtain a discrete joint density, obtaining an ``image'' where each pixel's brightness equals the normalised density of samples in the corresponding bin.
The resulting matrix $T_r \in [0,1]^{10 \times 10}$ serves as the \emph{reference template}, which represents the activity carried out at startup or when a transition is detected. Thus, the reference template is updated at every activity change recognised by the gate.

\subsection{Current Template Creation}

Although the \emph{reference template} is computed only when a change has been detected, another template, namely the \emph{current template}, is computed for each new window.
At every window boundary, a \emph{current template} $T_n \in [0,1]^{10 \times 10}$ is built from the most recent $W$ samples using the same two features selected in phase~(B), chosen among the features computed for each window in phase~(A).

The two features are normalised through the values $\min(f_k)$, $\max(f_k)$ saved from the reference window, so that both templates share the same bin boundaries. Bin counts are then normalised by the window size and accumulated into a new $10 \times 10$ histogram following~(\ref{eq:rescaling}) and providing the \emph{current template}.

\subsection{Comparison and Decision}
\label{sec:comparison}
The reference and current template can be directly compared, as their information content differs, but their underlying structure is equal. 
If they are deemed similar enough, no change is detected, and a new window can be analysed. Otherwise, the presence of a transition is identified. 

Given $T_r$ and $T_n$, the similarity is quantified by the Normalised Correlation Coefficient~(NCC), computed over all $P = 100$ bin
indices $p$:
\begin{equation}
    NCC = \frac{\displaystyle\sum_p \bigl(T_n[p] - \bar{T}_n\bigr)\bigl(T_r[p] - \bar{T}_r\bigr)}
    {\sqrt{\displaystyle\sum_{p} \bigl(T_n[p] - \bar{T}_n\bigr)^2} \cdot
     \sqrt{\displaystyle\sum_{p} \bigl(T_r[p] - \bar{T}_r\bigr)^2}},
  \label{eq:NCC_computation}
\end{equation}
where $\bar{T}_n$ and $\bar{T}_r$ are the respective template means.
A value of $NCC = 1$ indicates a perfect match; values near zero or negative indicate a substantial dissimilarity.

\begin{figure}[b!]
      \centering
      \includegraphics[width=1\linewidth]{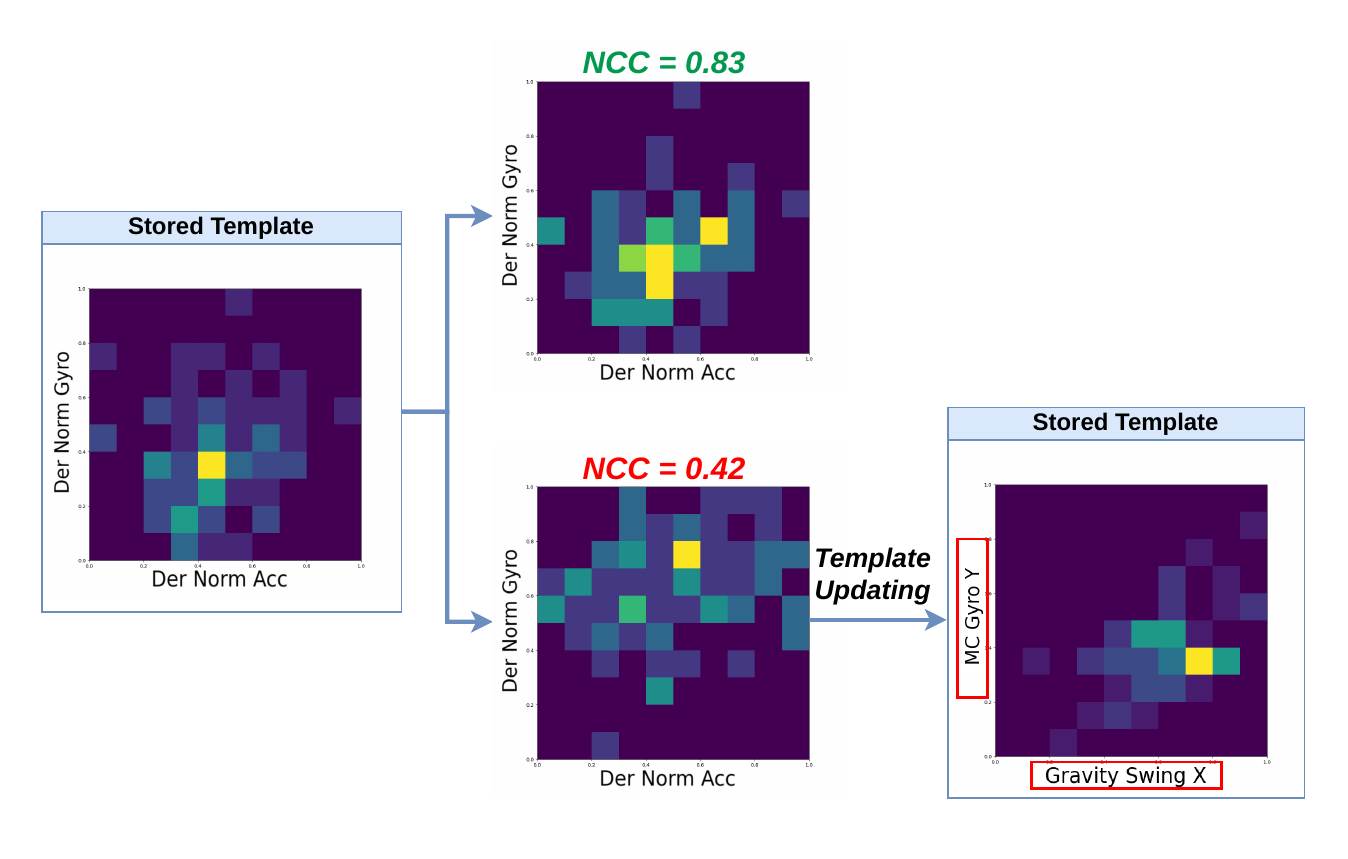}
      \caption{Template comparison: high NCC indicates no change; low NCC triggers a transition detection and template update.}
      \label{fig:templ_updating}
\end{figure}

As shown in Fig.~\ref{fig:templ_updating}, the NCC is then compared against a fixed threshold $Th_{NCC}$, computed as shown in Section \ref{sec:calibration}. When $NCC \geq Th_{NCC}$, the activity is stable, and no transition is detected; otherwise, a transition is identified, the HAR network is invoked to perform classification, and phase~(B) is rerun to obtain a new $T_r$. 

\subsection{Threshold Calibration}
\label{sec:calibration}

\begin{figure}[b!]
    \centering
    \includegraphics[width=1\linewidth]{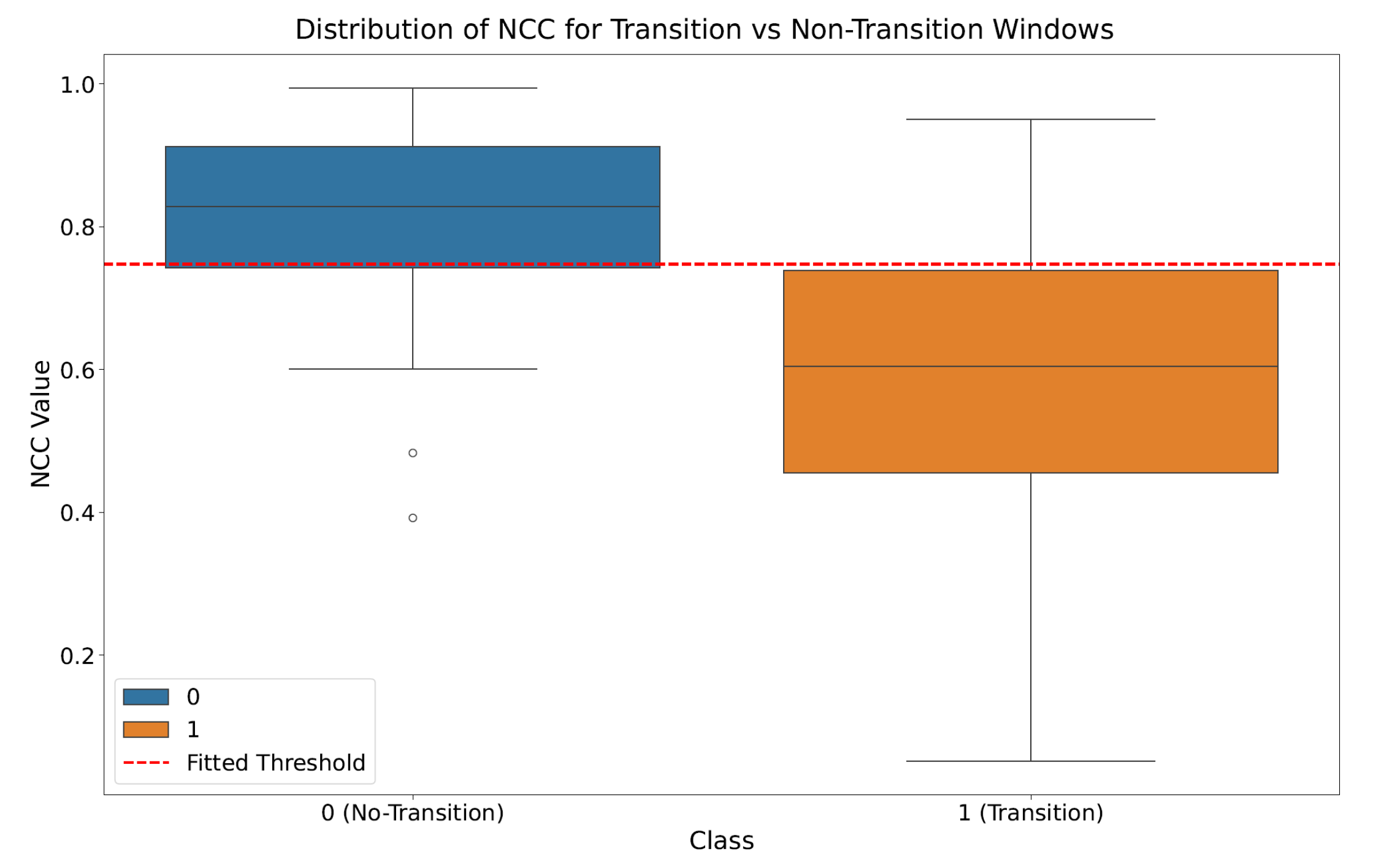}
    \caption{Example of a boxplot for the distribution of NCC values for the classes 0 (no transition) and 1 (transition) retrieved during the calibration phase. The red line represents the optimal fit of the NCC value for the considered dataset.}
    \label{fig:boxplots}
\end{figure}

To further tailor the sensitivity of the algorithm for a coarser or finer identification of changes, the threshold $Th_{NCC}$ is selected once via a short supervised calibration phase. This phase emulates the inference process while already knowing the underlying samples' labels. The update of $T_r$ is forced in correspondence with known transitions and the NCC values between all consecutive templates are saved. The resulting NCC time series is then used to find the value of $Th_{NCC}$ that maximises a weighted combination of True Positive Rate (TPR) and True Negative Rate (TNR), with a higher weight placed on TPR to prioritise reliable transition detection. The selected threshold optimally separates Class~1 (windows with a transition) from Class~0 (windows without a transition), as shown in Fig.~\ref{fig:boxplots}. As this procedure stores only the scalar NCC values and no raw sensor data, it is compatible with future on-device execution, guaranteeing user or device personalisation.

An explorative study to assess the minimum calibration size shows that the use of just 2 to 3 minutes of recorded activities with 8 to 10 transitions can be sufficient to efficiently retrieve the optimal threshold. 

\section{Experimental setup}
\label{sec:experiments}

\subsection{Datasets}
The gate algorithm is evaluated on two publicly available IMU datasets spanning three device types. The only data-dependent step differing between the datasets is the threshold calibration described in Subsection ~\ref{sec:calibration}.

\paragraph{UCA-EHAR~\cite{Novac2022}}
Data from 20 subjects at 26\,Hz, IMU on smart glasses, six protocols, spanning ten activity classes:
\textsc{lying}, \textsc{running}, \textsc{sitting}, \textsc{stairs},
\textsc{standing}, \textsc{walking}, \textsc{stand to sit}, \textsc{sit to stand}, \textsc{sit to lie}, \textsc{lie to sit} (\textsc{drinking} discarded as
fine-grained).

\paragraph{WISDM~\cite{Weiss2019}}
Data from 51 subjects at 20\,Hz, 6-axis IMU simultaneously on smartwatch and
smartphone, five activities: \textsc{jogging}, \textsc{sitting},
\textsc{stairs}, \textsc{standing}, \textsc{walking} (other fine-grained activities are discarded).

\subsection{Preprocessing}

Feature extraction is performed using a 3\,s sliding window with 50\% overlap, resulting in windows of $W$ samples and improving transition detection. This choice is consistent with prior HAR studies, which report that 2.5--3.5\,s windows are optimal for activity classification~\cite{Wang2018}.

\begin{figure}[b!]
    \centering
    \includegraphics[width=0.85\linewidth]{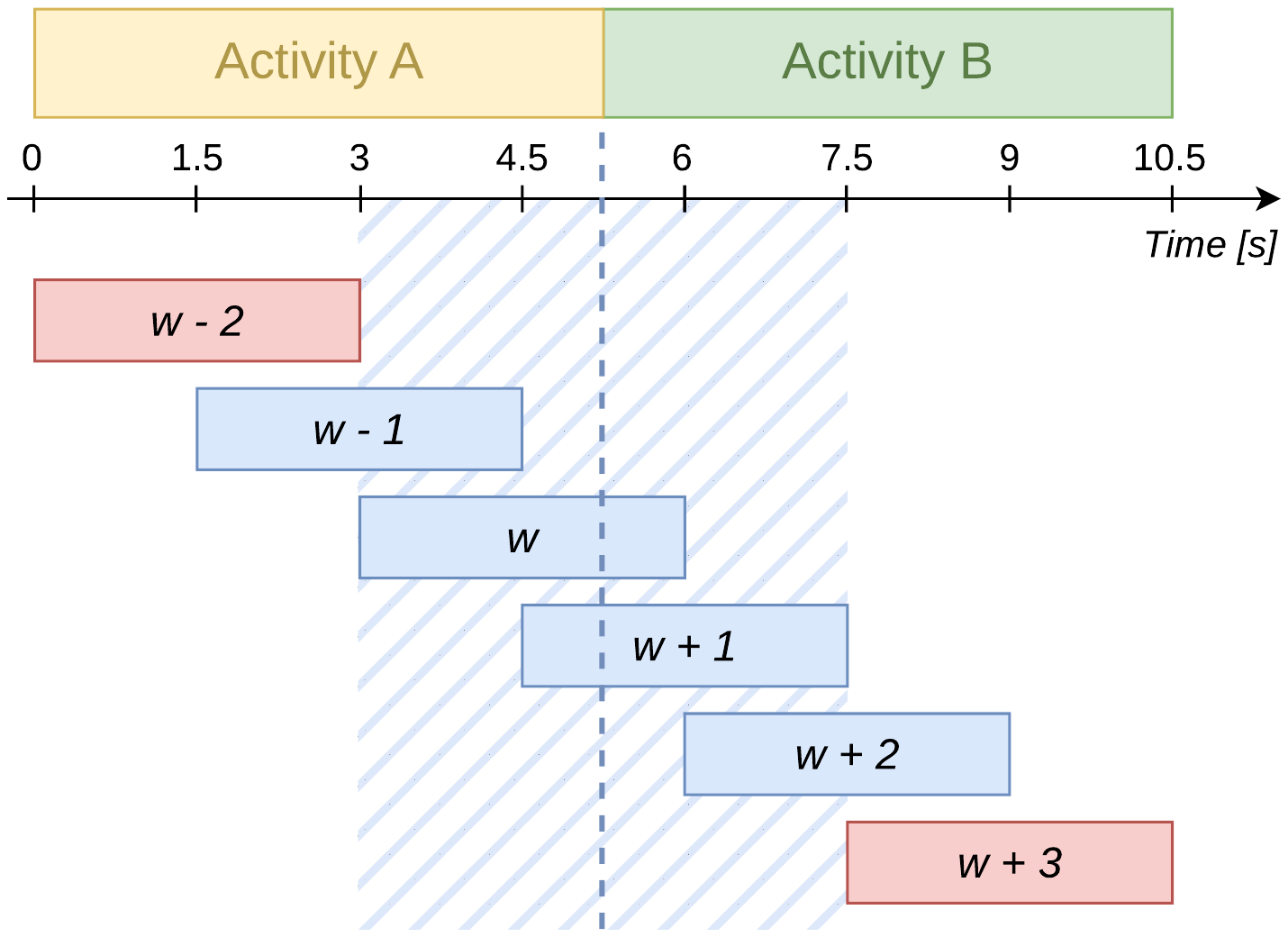}
    \caption{Windowing with 50\% overlap. Detection in the blue windows is correct; red windows yield False Positives.}
    \label{fig:windowing}
\end{figure}

As transitions near the boundaries of a window may fail to cause a significant change in the template, a relaxed evaluation criterion is adopted.
As shown in Fig.~\ref{fig:windowing}, a detection is considered correct if it falls in any of the windows spanning from $(w-1)$ to $(w+2)$, allowing a latency margin of $\pm 3$\,s relative to the ground-truth change point.

\subsection{Baselines}
\label{sec:baselines}

Three baseline approaches were implemented to benchmark the proposed algorithm: the CUSUM algorithm, a Decision Tree (DT), and a Support Vector Machine (SVM).
The CUSUM algorithm does not require a training phase, but relies on three tuneable parameters: sensitivity, warm-up samples, and detection threshold~\cite{Buroni2024}.
These were empirically tuned by iteratively varying their values and selecting the combination that maximised the G-mean score.

The DT and SVM recast the task as binary classification, labelling each window as containing a transition (Class~1) or not (Class~0).
The same 12 features used by the proposed algorithm were extracted window-wise and normalised to $[0,\,1]$. To address the severe class imbalance in the training set, both models were trained on both the original distribution and a balanced variant obtained by downsampling the majority class and partially duplicating the minority class.
Hyperparameter optimisation was performed via grid-search cross-validation, using recall as the scoring metric. The DT search space included the split criterion (\{Gini, Entropy\}), maximum depth (\{10, 20, 30\}), minimum samples per leaf (\{1, 2, 4\}), and minimum samples to split (\{2, 4\}). For the SVM, the kernel type (\{Linear, RBF\}), regularisation parameter (\{1, 5, 10, 20\}), and kernel coefficient (\{0.1, 1, 5\}) were explored. All ML results were averaged over 10 independent runs with different random seeds to account for partition sensitivity.

\section{Experimental results}
\label{sec:results}

\begin{table}[b!]
    \centering
    \caption{Optimal parameters for the baseline methods.}
    \label{tab:optimal_params}
    \begin{tabular}{l l l}
        \toprule
        \textbf{Model} & \textbf{Parameter} & \textbf{Optimal Value} \\
        \midrule
        \multirow{3}{*}{CUSUM \cite{Buroni2024}}          & \texttt{sensitivity}          & 3           \\
                                        & \texttt{warm-up samples}      & 26 (1\,s)   \\
                                        & \texttt{threshold}            & 0.3         \\
        \midrule
        \multirow{4}{*}{Decision Tree}  & \texttt{criterion}            & Entropy     \\
                                        & \texttt{max\_depth}           & 20          \\
                                        & \texttt{min\_samples\_leaf}   & 4           \\
                                        & \texttt{min\_samples\_split}  & 2           \\
        \midrule
        \multirow{3}{*}{SVM}            & \texttt{kernel}               & RBF         \\
                                        & \texttt{regularization paramter}                    & 10          \\
                                        & \texttt{gamma}                & 5           \\
        \bottomrule
    \end{tabular}
\end{table}

\subsection{Detection Performance and Comparison}
\label{sec:detection_performance}
The proposed gate is compared with a CPD method, i.e., the \textit{CUSUM} algorithm \cite{Buroni2024}, and with the supervised baselines introduced in Section~\ref{sec:baselines}, all evaluated using the optimal hyperparameters reported in Table~\ref{tab:optimal_params}. The comparison is conducted across all three device configurations under the relaxed evaluation protocol detailed in Subsection~\ref{sec:comparison}, achieving the results illustrated in Table~\ref{tab:results_all}. 

The algorithm's overall sensitivity sits at 98\% for the UCA-EHAR dataset \cite{Novac2022}, whereas the sensitivity is around 75\%.
On the WISDM dataset \cite{Weiss2019}, the approach achieves a TPR of approximately 97\% and a TNR of 76\%, both for the smartphone and the smartwatch sets, thus demonstrating its generalisation capabilities across heterogeneous sensors. 
The achieved results reflect the design choice of sensitivity maximisation, despite a decrease in specificity: this is deliberately intended to ensure that all changes are identified, at the expense of more frequent activations of the MCU.

The \textit{CUSUM} algorithm \cite{Buroni2024} yields a TPR between 91\% and 94\%, but demonstrates poor generalisation on diverse change types, due to the manual choice of specific parameters that fall short in detecting both dynamic and static transitions. 

Standard supervised methods reach inconsistent performances across different devices. The TNR ranges from 91\% to 99\%, proving good robustness to false positives. However, the maximum TPR for the UCA-EHAR dataset \cite{Novac2022} is below 91\%, and significantly drops for the WISDM dataset \cite{Weiss2019} (reaching 79\% for the smartwatch set and 61\% for the smartphone set).

Specifically, when employing an unbalanced training set, which is representative of realistic monitoring settings, the models perform poorly; when balancing the training set, the TPR increases. However, the results are shown to depend on the specific splitting of the validation set, as demonstrated by the higher standard deviation values compared to the proposed algorithm. Moreover, these models rely on an extensive training dataset to perform effective inference.

\begin{table*}[t!]
\caption{Detection performance and comparison across datasets (mean $\pm$ std across subjects). "Unbal." = unbalanced training set; "Bal." = balanced training set.}
\label{tab:results_all}
\begin{center}
\renewcommand{\arraystretch}{1.15}
\footnotesize
\begin{tabular}{llccc}
\toprule
\textbf{Dataset} & \textbf{Method} & \textbf{TPR [\%]} & \textbf{TNR [\%]} & \textbf{Training} \\
\midrule
\multirow{6}{*}{\shortstack[l]{\textbf{UCA-EHAR}\\(smart glasses)}}
  & CUSUM~\cite{Buroni2024} & 93.97${\pm}$3.53 & 78.78${\pm}$2.35 & Param.\ fit \\
  & DT (Unbal.)             & 86.85${\pm}$9.43 & 93.66${\pm}$1.99 & Offline \\
  & DT (Bal.)               & 91.07${\pm}$12.35 & 91.19${\pm}$3.30 & Offline \\
  & SVM (Unbal.)            & 85.23${\pm}$5.54 & 97.24${\pm}$1.64 & Offline \\
  & SVM (Bal.)              & 91.16${\pm}$8.17 & 93.50${\pm}$2.40 & Offline \\
  & \textbf{Proposed}       & \textbf{97.58${\pm}$2.34} & 74.71${\pm}$1.56           & None \\
\midrule
\multirow{4}{*}{\shortstack[l]{\textbf{WISDM}\\(smartwatch)}}
  & CUSUM~\cite{Buroni2024} & 93.00${\pm}$9.54   & 78.43${\pm}$5.09         & Param.\ fit \\
  & DT (Bal.)               & 79.11${\pm}$21.53       & 96.44${\pm}$2.10        & Offline \\
  & SVM (Bal.)              & 68.82${\pm}$17.76       & 98.27${\pm}$1.13         & Offline \\
  & \textbf{Proposed}       & \textbf{96.73${\pm}$6.76}         & 75.26${\pm}$2.42         & None \\
\midrule
\multirow{4}{*}{\shortstack[l]{\textbf{WISDM}\\(smartphone)}}
  & CUSUM~\cite{Buroni2024} & 91.17${\pm}$11.97         & 72.86${\pm}$5.81         & Param.\ fit \\
  & DT (Bal.)               & 58.25${\pm}$17.94         & 97.96${\pm}$1.71         & Offline \\
  & SVM (Bal.)              & 60.75${\pm}$17.14          & 98.99${\pm}$1.03         & Offline \\
  & \textbf{Proposed}       & \textbf{97.34${\pm}$5.42}       & 75.92${\pm}$2.50       & None \\
\bottomrule
\end{tabular}
\end{center}
\end{table*}

\subsection{Computational Cost and Savings}
\label{sec:complexity}

The proposed change detection method was assessed for future on-device development.
The algorithm's memory allocation requirement mainly revolves around the storage of the floating-point feature values.
Considering 3-second windows at 26 Hz, as for UCA-EHAR~\cite{Novac2022}, the estimated memory allocation for these features is 4.53 kB. It is of note that this value is low enough to allow the algorithm to be embedded not only in MCUs, but also in smart sensor units~\cite{STMicroelectronics_ISPU}. 

For what concerns the computational load, the gate costs approximately 16\,kFLOPs per step, making it more than 50$\times$ cheaper than a HAR network such as Dendron~\cite{11007628}, with approximately $853$\,kFLOPs.

\begin{figure}[t!]
    \centering
    \includegraphics[width=1\linewidth]{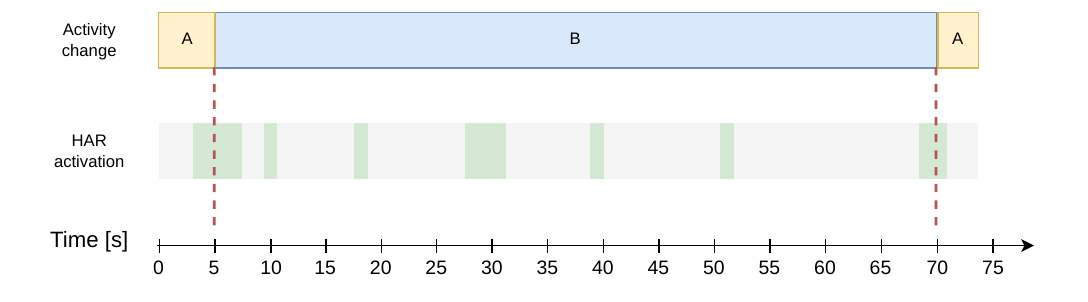}
    \caption{Example of the proposed methodology on subject T15 from the UCA-EHAR dataset \cite{Novac2022} during a running session. The top bar shows the ground-truth activity sequence, with transitions from \textsc{standing} to \textsc{running} and back to \textsc{standing}; red dashed lines mark the corresponding change points. In the lower bar, green is where the gate triggers the HAR pipeline, whereas gray indicate no activation.}
    \label{fig:HAR_activation}
\end{figure}

This clearly shows the advantages of using a lightweight gate to activate a HAR algorithm. Specifically, taking UCA-EHAR~\cite{Novac2022} as an example, the HAR network would be invoked only on 30\% of the windows (914 out of 2966), including those identified as false positives by the gate.
By considering the computational load of the gate and a network such as Dendron~\cite{11007628}, we obtain that our proposed framework, i.e., keeping the gate always on, and invoking HAR when a change is detected, would lead to a reduction equal to about 67\% in terms of number of FLOPs.
A graphical example of the reduced number of windows for which the HAR network is run is presented in Fig.~\ref{fig:HAR_activation}.



\section{Conclusions and Future Developments}
The proposed framework provides a lightweight solution for online change-point detection in on-device HAR. By processing samples sequentially as they are acquired, the method operates in real time without storing raw windowed data, making it well-suited to extreme-edge deployment. 

Unlike conventional ML and DL approaches, it does not rely on large training datasets, model compression, or quantization, and requires only a brief device-specific calibration phase. This low-overhead design also makes a fully on-sensor implementation feasible.

Experimental results on UCA-EHAR~\cite{Novac2022} and WISDM~\cite{Weiss2019} show that the method generalizes across heterogeneous devices, including smart glasses, a smartwatch, and a smartphone. The gate achieves high sensitivity, with average values of approximately 97\% on UCA-EHAR and WISDM, ensuring that nearly all activity transitions trigger an HAR update. At the same time, specificity reaches 75\% and 76\%, respectively, reducing unnecessary invocations while preserving responsiveness to underlying activity changes. Compared with existing approaches, the method offers a favourable trade-off between robustness, simplicity, and deployment cost.

Beyond real-time inference, the main practical advantage of the proposed approach lies in enabling change-driven activation of the MCU, which could substantially reduce power consumption by limiting both on-board processing and serial communication to sparse, informative feature updates. 

Although this work yielded promising results, further developments are needed to assess the method under a broader and more structured evaluation protocol, and to validate it through direct deployment on physical devices in realistic, real-time monitoring conditions.

\section{ACKNOWLEDGMENTS}

This work was carried out in the EssilorLuxottica Smart Eyewear Lab, a Joint Research Center between EssilorLuxottica and Politecnico di Milano.

{\small
\bibliographystyle{ieeetr}
\bibliography{egbib}

}

\end{document}